\documentclass[12pt]{article}

\usepackage{epsfig}
\usepackage[usenames]{color}

\def\be{\begin{equation}}
\def\ee{\end{equation}}
\newcommand{\bea}{\begin{eqnarray}}
\newcommand{\eea}{\end{eqnarray}}
\def\bd{\begin{displaymath}}
\def\ed{\end{displaymath}}

\def\ba{\begin{eqnarray}}
\def\ea{\end{eqnarray}}

\definecolor{red}{rgb}{1,0,0}

\begin{document}

\vspace{18pt} \vskip 0.01in \hfill TAUP-287908 \vskip 0.01in \hfill {\tt
hep-th/yymmnnn}

\vspace{30pt}

\begin{center}
{\bf \LARGE  On the spectra of scalar mesons from HQCD models}
\end{center}

\vspace{30pt}

\begin{center}
Oded Mintakevich and  Jacob Sonnenschein

\vspace{20pt}

\textit{School of Physics and Astronomy\\ The Raymond and Beverly
Sackler Faculty of Exact Sciences\\ Tel Aviv University, Ramat Aviv
69978,
Israel\\[10pt]
 }

\end{center}


\begin{center}
\textbf{Abstract }
\end{center}
We determine the  holographic spectra of scalar mesons from the
fluctuations of the embedding  of flavor D-brane  probes in  HQCD
models. The models we consider include a generalization of the Sakai
Sugimoto model at zero temperature and  at the ``high-temperature
intermediate phase", where the system is in a deconfining phase
while admitting chiral symmetry breaking and a non-critical 6d model
at zero temperature. All these models are based on backgrounds
associated with near extremal $N_c$ D4 branes and a set of
$N_f<<N_c$ flavor  probe branes that admit geometrical chiral
symmetry breaking. We point out that the spectra of these models
include a $0^{--}$ branch   which does not show up in nature.
 At zero temperature we found that the masses of
the mesons $M_n$ depend on the ``constituent quark mass"
parameter $m^c_q$ and
on the excitation number $n$ as  $M_n^2\sim m^c_q$ and $M^2_n \sim
n^{1.7}$  for the ten dimensional case and as $M_n\sim m^c_q$ and
$M_n \sim  n^{0.75}$ in the non-critical case. At the high temperature
intermediate phase we detect a decrease of the masses of low spin
mesons as a function of the temperature similar to holographic
vector mesons and to lattice calculations.


\vspace{4pt} {\small \noindent

 }  \vfill
\vskip 5.mm
 \hrule width 5.cm
\vskip 2.mm {\small \noindent E-mails :odedm@post.tau.ac.il
cobi@post.tau.ac.il}

\thispagestyle{empty}

\eject

\setcounter{page}{1}

 Whereas realizing confinement in dual holographic models of QCD  (HQCD)
is easy the incorporation of  flavored chiral quarks and in
particular chiral symmetry breaking is more difficult. Sakai and
Sugimoto \cite{Sakai:2004gf} proposed   a model that admits the two
phenomena. It is based on placing a set of $N_f$ D8 and anti D8
probe flavor branes into the gravity model of near extremal D4
branes \cite{Witten:1998zw,Itzhaki:1998dd}.

The mesonic spectra is one of the most important properties of
hadron dynamics that  can  be  ``measured"  in the HQCD  laboratory.
The low spin mesons are associate with the fluctuations of the fields
that reside on the probe flavor branes, the vector mesons with the
$U(N_f)$ flavor gauge fields and the scalar mesons with the
embedding of the  probe branes
 \footnote{  High spin mesons are naturally described  by
 semi-classical spinning string
configurations\cite{Kruczenski:2004me}}. Here in this paper we focus
only  on  scalar mesons. The motivation behind  addressing this
problem are the following:
 (i) To verify that the meson spectrum at zero temperature does not include tachyonic modes. Had there been such modes it would have indicated that the system is unstable. Since the model of \cite{Sakai:2004gf}
is based on placing branes and anti-branes one may be worried that
the system  is unstable and hence the importance of this check. (ii)
The spectrum of the scalar mesons has been  determined already in
\cite{Sakai:2004gf}. However the attemps to derive it in
generalizations of the model where the asymptotic separation of the
brane anti-brane $L$ is smaller than half of the circumference of
the compactified direction $x_4$, namely for $L\leq \pi R$ failed
for the symmetric modes \cite{Casero:2005se,Peeters:2006iu}
(iii) To determine the dependence of the spectrum on the excitation
number $n$ and the parameter $m^c_q$  defined in
(\ref{constituentmass}) that is related to the constituent quark
mass. In addition  one naturally would like to compare the explicit
ratios of meson masses that one deduces from any given  HQCD model
and the experimental data to get an indication of how well the model
describes real hadron physics. (iv) To further examine the
differences of physical properties extracted from critical models to
non-critical models which were previously discussed in
\cite{Casero:2005se,Kuperstein:2004yf,Mazu:2007tp}. The spectrum of
scalar mesons was extracted also in other HQCD models\cite{Erdmenger:2004dk,
Hovdebo:2005hm,Antonyan:2006pg}. For further
reading see \cite{Erdmenger:2007cm} and references therein.

We can summarize the outcome of the paper as follows
\begin{itemize}
\item
We were able to choose coordinates that avoid the singularities that
were encountered in previous works \cite{Casero:2005se,Peeters:2006iu} and determine the spectrum of both the
anti-symmetric as well as symmetric branches.
\item
We find that in the models examined and in particular the original
model of \cite{Sakai:2004gf} the symmetric solutions correspond to
scalar mesons of the form $0^{++}$ whereas the anti-symmetric
solutions correspond to $0^{--}$ mesons. This property which seems
to be in common to a HQCD models based on probe branes and
anti-branes, contradict the low lying spectrum in nature. There are
no low lying $0^{--}$ mesons.

\item
At zero temperature we found that the masses of the mesons $M_m$
depend on the ``constituent quark mass" $m^c_q$  and on the excitation
number $n$ as  $M_m^2\sim m^c_q$ and $M^2_m \sim n^\alpha$ with
$\alpha\sim 1.7$ for the ten dimensional case and as $M_m\sim m^c_q$
and $M_m \sim  n^\beta$  with $\beta\sim 0.75$ when a CS term is incorporated
 and  $\beta\sim 1$ without such a term  in the non-critical
case. At the high temperature intermediate phase we detect a
decrease of the masses  of low spin mesons as a function of the
temperature similar to holographic vector mesons and to lattice
calculations.
\end{itemize}

The paper is organized as follows. We begin in section \ref{section1} with a
brief review of the holographic models we investigate. We summarize
the main features of the model of Sakai and Sugimoto at zero and
finite temperature and an analogous six dimensional non-critical
model. In section \ref{dvir} we describe the extraction of scalar mesons from
the fluctuations of the embedding. In particular we point out that
in the coordinates introduced in \cite{Sakai:2004gf} the eigenvalue
problem admits  a singularity that prevents the numerical
determination of the eigenvalues. A different coordinate system is
presented in section \ref{section3} which evades the problem of the singularity.
Using these coordinates, the spectrum of masses as a function of the
constituent mass and excitation number is derived.
The spectrum of scalar mesons that follows from  a non critical model of near extremal D4 branes is analyzed in section \ref{non_crit}. Section \ref{parity}
addresses the issue of parity and charge conjugation of the
scalar mesons. It is pointed out that the spectrum includes $0^{--}$
mesons which do not show up in nature.
Section \ref{intermediate} is devoted to the spectrum of mesons above the deconfining phase transition
in the ``intermediate phase".
We summarize the results and raise certain
open questions.


\section{Review of the holographic models}\label{section1}
\subsection{The Sakai Sugimoto model}
The model of \cite{Witten:1998zw}, describes the near horizon limit of
$N_c$ $D4$-branes wrapping a circle in the $x_4$ direction with anti periodic
boundary condition for the fermions.
Into this background a stack of $N_f \ D8$ is placed
at $x_4=0$ and a stack of $N_f \  \bar{D}8$ at the
anti-podal point of the $x_4$ circle \cite{Sakai:2004gf}.
Assuming $N_f<<N_c$ one can overlook the modification of the
metric and dilaton due to the backreaction of the background
by the $N_f$ $D8$-$\bar{D}8$ systems and continue to use the
metric and dilaton associated with the $N_c$ $D4$ alone.
Therefore the metric, dilaton and the RR four form are given by
\begin{eqnarray}\label{S_S_metric}
ds^2\!\!\!\!&=&\!\!\!\bigg( \frac{u}{R_{D4}}\bigg)^{3/2}\!\bigg[\! \!-\!\!dt^2\!+\!\delta_{ij}dx^idx^j+f(u)dx_4^2\bigg]
\!+\!\bigg( \frac{R_{D4}}{u}\bigg)^{3/2}\!\bigg[\frac{du^2}{f(u)}\!+\!u^2d\Omega_4^2\bigg]\\ \nonumber
F_4\!&=&\!\frac{2\pi N_c}{V_4}\epsilon_4\ \ ,\ \ e^{\phi}=g_s\bigg( \frac{u}{R_{D4}}\bigg)^{3/4}
 , \ R_{D4}^3=\pi g_sN_cl_s^3 \ , \  \ f(u)=1-\bigg( \frac{u_{\Lambda}}{u}\bigg)^3
\end{eqnarray}
Where $V_4$ denotes the volume of the unit sphere $\Omega_4$ and $\epsilon_4$ its corresponding
volume form. $l_s$ is the string length and $g_s$ a parameter related to the string coupling.
The $x_4$ is the compactified direction that is asymptotically transverse to the $D8$. The manifold spanned by the coordinate $u,x_4$ has the topology of a cigar where its tip is at the minimum value of $u$ which is $u=u_{\Lambda}$.
The periodicity of this cycle is uniquely determine to be
\begin{eqnarray}
\delta x_{4}=\frac{4\pi}{3}\bigg( \frac{R_{D4}^3}{u_{\Lambda}}\bigg)^{1/2}=2\pi R
\end{eqnarray}
in order to avoid a conical singularity at the tip of the cigar.
The classical profile of the $D8$ probe brane in this background
is given by the classical solution to the e.o.m of the DBI action
of that probe brane. The $D8$ DBI action is
\begin{eqnarray}\label{S_D8}
&S_{D8}&=T_8\int dtd^3xdud\Omega_4e^{-\phi}\sqrt{-det\hat{g}}=
\tilde{T}_8 \int dtd^3xduu^4\sqrt{f(u)(\partial_ux_4)^2+
\frac{R^3_{D4}}{u^3f(u)}}\\
&=&\tilde{T}_8 \int dtd^3xdx_4u^4\sqrt{f(u)+
\bigg( \frac{R_{D4}}{u}\bigg)^3\frac{u'^2}{f(u)}}
\end{eqnarray}
where $\hat{g}$ stands for the pullback metric on the $D8$ brane.
The simplest way of solving this e.o.m is by noting that the action is independent of
$x_4$ and so its Hamiltonian is conserved.
\begin{equation}
\frac{u^4f(u)}{\sqrt{f(u)+\bigg( \frac{R_{D4}}{u}\bigg)^3\frac{u'^2}{f(u)}}}=u_0^4\sqrt{f(u_0)}
=const
\end{equation}
where we assumed that there is a point $u_0$ where the curve $u(x_4)$, which
describes the profile of the $D8$ brane in the $(u,x_4)$ plane has a minimum.
At that point the $D8$ brane bends, namely the $D8$-$\bar{D}8$ join
together. After some algebra one finds
\begin{eqnarray}\label{partial_x_4}
\bigg(\frac{\partial x_{4}}{\partial u}\bigg)_{cl}=
\frac{1}{f(u)(\frac{u}{R_{D4}})^{3/2}\sqrt{\frac{f(u)u^8}{f(u_0)u_0^8}-1}}
\end{eqnarray}
Hence we find that the profile of the $D8$ brane probe is
\begin{eqnarray}\label{u_Lamabda_ne_u}
x_{4}(u)=\int^{u}_{u_0}
\frac{du}{f(u)(\frac{u}{R_{D4}})^{3/2}\sqrt{\frac{f(u)u^8}{f(u_0)u_0^8}-1}}
\end{eqnarray}
Where $u_0$ is a constant of integration setting the lowest value of $u$
to which the $D8$ barne is extending. At that point the
$D8$ brane join the $\bar{D}8$ brane and the brane is extending back into the UV.
The value of $u_0$ also sets the asymptotic distance $L$ between the
position of the $D8$ and $\bar{D}8$ brane
\begin{eqnarray}\label{L}
L=\int dx_4=2\int_{u_0}^{\infty}\frac{du}{u'}=
2(\frac{R_{D4}^3}{u_0}\bigg)^{1/2}\int_1^{\infty}dy
\frac{y^{-3/2}}{\sqrt{f(y)}\sqrt{\frac{f(y)}{f(1)}y^8-1}}
\end{eqnarray}
Hence we see
\begin{eqnarray}\label{L_u_0}
L\propto \bigg(\frac{R_{D4}^3}{u_0}\bigg)^{1/2}
\end{eqnarray}
For later use we define
\begin{eqnarray}\label{gamma}
\gamma=\frac{u^8}{f(u)u^8-f(u_0)u_0^8}
\end{eqnarray}
The DBI action then becomes
\begin{eqnarray}\nonumber
S=T_8\int e^{-\phi}\sqrt{|det\hat{g}_0|}\sim \int d^4xdu\gamma^{1/2}u^{5/2}
\end{eqnarray}

\subsection{Thermodynamics of the Sakai Sugimito model}\label{termo}
In \cite{Aharony:2006da} \footnote{see also \cite{Parnachev:2006dn}.} a study of the thermodynamics of the
Sakai Sugimito model was carried using the conjecture presented
in \cite{Witten:1998zw}.\\
The conjecture states that the thermodynamics
of a field theory with a gravitational dual
is determined by taking into account the contribution to the saddle point approximation
from all the gravitational backgrounds with the correct 'UV' asymptotic,
with compactified Euclidean time direction of period $\beta=\frac{1}{T}$ and with anti-periodic boundary condition for the fermions along this direction.
The temperature of the field theory is $T=1/\beta$
and its properties are read from
the manifold responsible for the most dominant contribution to the
saddle point approximation namely the one that has the
lowest free energy.\\
When ever one background looses its domination to another  background
as we vary the temperature, a phase transition occurs in the dual
field theory.

In \cite{Aharony:2006da} two manifolds where found to have the same 'UV'
asymptotic as the one of Sakai and Sugimoto model,
the background (\ref{S_S_metric}),
and the same configuration only with the time and $x_4$ directions interchange.
\begin{eqnarray}\label{High_T_S_S_metric}
ds^2\!\!\!\!\!&=&\!\!\!\!\!\bigg( \!\frac{u}{R_{D4}}\!\bigg)^{3/2}\!\!\!\!\!\!\![-f(u)dt^2\!\!+\delta_{ij}dx^idx^j+dx_4^2]
+\bigg(\! \frac{R_{D4}}{u}\!\bigg)^{3/2}\!\bigg[\frac{du^2}{f(u)}+u^2d\Omega_4^2\bigg]
\end{eqnarray}
with
\begin{eqnarray}
f(u)=1-\bigg(\frac{u_T}{u}\bigg)^3
\end{eqnarray}
and the temperature is given by
\begin{eqnarray}
\delta t=\frac{4\pi}{3}\bigg(\frac{R^3_{D4}}{u_T}\bigg)^{1/2}=\beta
\end{eqnarray}
The difference between the free energy densities of the two backgrounds is proportional
to $N_c^2[(2\pi T)^6-1/R^6]$ \footnote{
Of course in our model there are also $D8$ brane which their DBI action
will contribute to the total free energy of the configuration as well,
but this is sub-leading to the bulk action since the bulk action is of
order $N_c^2$ and the contribution of the $D8$ is of order $N_c\cdot N_f$
which is negligible in the probe approximation}.
This means that when the circumference of the $x_4$ cycle
is smaller than that of the time
direction namely when $T<1/2\pi R$
the background  (\ref{S_S_metric}) is the dominant one,
while when the opposite occurs and $T>2\pi R$ the action of
(\ref{High_T_S_S_metric}) will dominates.
At the temperature $T=1/2\pi R$ the two actions are the same
since the two backgrounds are different by the labeling of the coordinates,
so at $T=T_c=1/2\pi R$ the system has a first order phase transition.
In \cite{Aharony:2006da} it was argue that in the dual
field theory, the physical interpretation to this phase transition
is a transition from confined phase at $T<1/2\pi R$
to deconfined at $T>1/2\pi R$.
This can be seen via a computation of the quark anti-quark
potential \cite{Kinar:1998vq} in the two backgrounds. Another
indication to this interpretation is that the renormalized free energy of
the low temperature phase shows a $N_c^0$ behavior while that of the
high temperature phase shows a $N_c^2$ one.
Hence from now on we will denote $T_c=T_d$.\\
At the high temperature phase there is another possible classical solution to
the profile of the $D8$ brane  which is
a configuration with constant $x_4$ namely, $x_4(u)=0,L$.\footnote{This configuration
was not possible in the low temperature, but in the high temperature phase
the time circle shrink to zero at $u=u_{\Lambda}$ and so the $D8$ brane
can just smoothly end there.}\\
Now since the bulk free energy is the same for the two configurations
of the $D8$ branes, the difference of the free energy of the $D8$ probes
determines which
of the two configurations is the preferable one for a given temperature.
It turns out that the transition between the two configuration depends
on the parameter $y_T=\frac{u_0}{u_T}$, its value at the phase transition
turns out to be $y_T^c\sim 0.73572$. \\
Using eq.(\ref{L}) we find $L_c=0.751\bigg(\frac{R_{D4}^3}{u_0}\bigg)^{1/2}$,
hence at the critical point $y_T=y_T^c$ the critical temperature is set by
the asymptotic distance between the branes (setting $R_{D4}=1$)
\begin{eqnarray}
T_c=\frac{3}{4\pi}u_T^{1/2}=\frac{3}{4\pi}(y_T^cu_0)^{1/2}=0.154/L
\end{eqnarray}
The field theory sees this transition as chiral symmetry restoration
at high temperature, this interpretation is natural since the $D8$
branes are now disconnected and there is an $U(N_f)\times U(N_f)$
global symmetry.\\
Hence we will denote this critical temperature as $T_{\chi SB}$.
Note that this only happens at the high temperature phase so
there is still the condition $T_{\chi SB}=0.154/L > 1/2\pi R$.\\
So if $L>0.97R$, we find that $T_d$ is always higher
than $T_{\chi SB}$, and so deconfinement and chiral symmetry
restoration phase transition happen together.
We see that in this model $\chi SB$ and confinement appear independently
of one another as a result of the existence of the free parameter $L$
coming from the $5d$ nature of the field theory.

\subsection{Non critical holographic model}
A non critical model with a very similar properties to
Sakai-Sugimoto model was presented in \cite{Casero:2005se,Kuperstein:2004yf}
\footnote{For other non-critical SUGRA models with flavor see
 \cite{Klebanov:2004ya,Israel:2005zp,Gursoy:2007er,Murthy:2006xt,Bigazzi:2006ix}.},
this model consists of non-extremal configuration of $N_c$ $D4$
branes placed in a six dimension space-time with one of the $D4$
coordinates taken to be periodic with anti periodic boundary
condition for the fermions.

The metric, dilaton and RR six-form field take the form \cite{Kuperstein:2004yf}
\begin{eqnarray}\label{noncriticalbackground}
ds^2&=&\bigg(\frac{u}{R_{AdS}}\bigg)^2dx_{1,3}^2+
\bigg(\frac{R_{AdS}}{u}\bigg)^2\frac{du^2}{f(u)}+
\bigg(\frac{u}{R_{AdS}}\bigg)^2f(u)dx_4^2 \\ \nonumber
F_{(6)}&=&Q_c\bigg(\frac{u}{R_{AdS}}\bigg)^4dx_0\wedge
dx_1\wedge dx_2\wedge dx_3\wedge du\wedge dx_4 \\ \nonumber
e^{\phi}&=&\frac{2\sqrt{2}}{\sqrt{3}Q_c}\ \ \  \ ; \ \ \ \ R^2_{AdS}=\frac{15}{2}
\end{eqnarray}
with
\begin{eqnarray}
f(u)=1-\bigg( \frac{u_{\Lambda}}{u}\bigg)^5
\end{eqnarray}
and where $Q_c$ is proportional to  $N_c $, the number of color $D4$ branes.
In order to avoid conical singularity
the periodicity of the cycle of $x_4$ is set to
\begin{eqnarray}
x_4\sim x_4+\delta x_4\ \ \ \  \ ; \ \  \ \ \
\delta x_4=\frac{4\pi R^2_{AdS}}{5u_{\Lambda}}
\end{eqnarray}
Of course the curvature of order one of this background makes the
leading order supergravity an un justified approximation to
string theory on this background.
Nevertheless its believed that at least the extremal model
due to its symmetries, is indeed a good
background for the study of non-critical string theory \cite{Klebanov:2004ya}.
Now we place $N_f$ $D4$ branes which are transverse to the $S^1$ cycle
and extend up to infinity in the $u$ direction.
The properties of the four dimensional low energy effective field theory
living on the intersection of these color and flavor $D4$ is then seem to be
very similar to those found at the Sakai Sugimoto model.
Thus we would like to study its spectrum of scalar excitations
and check if there is no tachyon in the model.

Just like in the critical model the $D4$ brane may bend on
the $(u,x_4)$ cigar and in order to find its profile one must solve
the e.o.m of the $x_4 $ coordinate.
This e.o.m is derived from the action of the
flavor $D4$ brans namely the DBI action plus the CS term
which are are given by
\begin{eqnarray}
S_{D4}=-T_4\int d^5xe^{-\phi}\sqrt{-\det(\hat{g})}+T_4\tilde{a}\int P(C_{(5)})
\end{eqnarray}

Following similar steps to does taken in the previous section we find
\begin{eqnarray}
x_{4,cl}(u)=\int_{u_0}^u
\frac{(u_0^5f^{1/2}(u_0)-au_0^5+au'^5)du'}{(\frac{u'}{R_{AdS}})^2f(u')
\sqrt{u'^{10}f(u')-(u_0^5f^{1/2}(u_0)-au_0^5+au'^5)}}
\end{eqnarray}
where $a=\frac{2}{\sqrt{5}}$.
\section{Fluctuation of the embedding and scalar mesons.}\label{dvir}
We now turn our attention to the study of the fluctuation of the
$D8$ brane around its classical profile. As was mentioned in the
itroduction, one has a two fold interest in those fluctuations: (i)
They correspond to scalar mesons in the dual gauge theory. (ii)
Tachyonic modes of  the fluctuation signals an instability of the
system.

We start by expanding the $x_4$ coordinate around its classical value
and define the fluctuation $\xi(u,x^{\mu})$ as follows:

\begin{eqnarray}
x_4(u,x^{\mu})=x_4(u)_{cl}+\xi(u,x^{\mu})
\end{eqnarray}
Substituting this into the action (\ref{S_D8}) and expanding to
quadratic order in $\xi$ we find the following action for
the fluctuations

\begin{eqnarray}\label{DBIscalar}
S\propto \frac{1}{2}\int d^4x du\bigg{\{}u^{5/2}R_{D4}^3{\gamma^{-1/2}}
\eta^{\mu\nu}\partial_{\mu}\xi\partial_{\nu}\xi
+ {u^{11/2}}{\gamma^{-3/2}}(\partial_u\xi)^2 \bigg{\}}
\end{eqnarray}
where $\gamma$  is defined in (\ref{gamma}). We now introduce the following mode expansion
\begin{eqnarray}\label{mode_expansion}
\xi(u,x^{\mu})=\sum_{n=0}^{\infty}f_n(x^{\mu})\xi_n(u)
\end{eqnarray}

Using the symmetries along the $x^{\mu}$  directions we have
\begin{eqnarray}
\eta^{\mu\nu}\partial_{\mu}\partial_{\nu}f_n=-m_n^2f_n
\end{eqnarray}
The e.o.m for the $\xi_n$ modes reads
\begin{eqnarray}
\partial_u[(u^{11/2}\gamma^{-3/2})\partial_u\xi_n]=-m_n^2R_{D4}^3u^{5/2}\gamma^{-1/2}\xi_n
\end{eqnarray}
or in its canonical form
\begin{eqnarray}\label{ux_eq}
\bigg{\{}\partial_u^2+\bigg[(\frac{12}{u}-\frac{15}{2u^4})\gamma
-\frac{13}{2u}\bigg]\partial_u\bigg{\}}\xi_n
=-\frac{m_n^2R_{D4}^3\gamma}{u^3}\xi_n
\end{eqnarray}

For $u_0>>u_{\Lambda}$, $f(u)\to 1$, the e.o.m
simplifies and the qualitative behavior of $m_n$ can be determined by
using dimensional arguments \cite{Aharony:2006da}. Define the  dimensionless parameter
$v=\frac{u}{u_0}$ then for the limit $u_0>>u_{\Lambda}$ where $f\to 1$
\begin{eqnarray}
\gamma\to \frac{1}{1-\frac{1}{v^8}}
\end{eqnarray}
The e.o.m in terms of $v$ reads
\begin{eqnarray}
\partial_v(v^{11/2}\gamma^{-3/2})\partial_v\xi_n=-m_n^2\frac{R_{D4}^3}{u_0}v^{5/2}\gamma^{-1/2}\xi_n
\end{eqnarray}
Since the L.H.S is dimensionless so must be the R.H.S and hence
\begin{eqnarray}\label{m_vs_u_0}
m_n^2 \propto \frac{u_0}{R_{D4}^3}
\end{eqnarray}
Using the relation (\ref{L_u_0}) between $u_0$ and $L$ we find
\begin{eqnarray}\label{m_vs_L}
m_n \propto \frac{1}{L}
\end{eqnarray}
while the mass of the glueball is related to $m_{gb}\sim \frac{1}{R}$.
For the case $u_{\Lambda}=u_0,\ \  L=\pi R$
so the glueball and mesons masses have the same scale. However in the general case where
$u_0>u_{\Lambda}$ there are two different scales $m_n\sim\frac{1}{L} >\frac{1}{R}\sim m_{gb}$.

In order to find the exact spectrum of the eigenvalues of
(\ref{ux_eq}) one can use the shooting technic which is implemented
by demanding symmetric or anti-symmetric boundary condition to
$\xi_n$ at $u=u_0$ and integrating the equation up to the $u>>u_0$
region where the solution could be matched to its normalizable
asymptotic expansion. Of course this matching is only possible when
the correct eigenvalues are being used and so one shoots with
different eigenvalues until a matching is obtained . However there
is a problem with these coordinates at $u=u_0$ since,
$\frac{dx_{4,cl}}{du}|_{u=u_0}\to \infty$ ( see eq.
(\ref{partial_x_4})). An odd perturbation to the classical
configuration will cause no change in the shape of this singularity
but an even one will, and so will also  have a singular derivative.

This problem is reflected in the singularity of the e.o.m (\ref{ux_eq}) at $u\to u_0$.
To see this behavior explicitly
we change coordinate to a dimensionless parameter $z$ as follows
\begin{eqnarray}\label{defz}
u^3=u_{0}^3+u_{\Lambda}^3z^2
\end{eqnarray}
the eigenvalue problem (\ref{ux_eq}) then becomes
\begin{eqnarray}\label{original_eq}
\bigg{\{}\partial_z^2+\bigg[ \frac{5u_{\Lambda}^3z}{u_{0}^3+u_{\Lambda}^3z^2}-\frac{1}{z}
-\frac{\gamma'u_{\Lambda }^3z}{(u_{0}^3+u_{\Lambda}^3z^2)^{2/3}\gamma}\bigg] \partial_z\bigg{\}}\xi_n
=-\frac{m_n^2R^3_{D4}u_{\Lambda }^6\gamma z^2}{(u_{0}^3+u_{\Lambda}^3z^2)^{7/3}}\xi_n
\end{eqnarray}
where $\gamma'$ stands for the derivative of $\gamma$ with respect to $u$. Since
\begin{eqnarray}
\gamma_{z\to 0}=\frac{\frac{3u_0^6}{u_{\Lambda}^3(8u_0^3-5u_{\Lambda}^3)}}{z^2}\ \ \ ; \ \  \
\gamma'_{z\to 0}=-\frac{\frac{9u_0^8}{u_{\Lambda}^6(8u_0^3-5u_{\Lambda}^3)}}{z^4}
\end{eqnarray}
we find that this equation has a regular singularity at $z=0$!\\
Indeed it was already noticed in \cite{Peeters:2006iu} that by
employing the 'shooting' technic only half of the spectrum could
be found, namely only the odd modes where seen
while the even ones could not be obtained, these modes that should
had been obtained by setting the boundary conditions to
\begin{eqnarray}
\xi_n(z=0)=1\ \ \ ;\ \ \ \partial_z\xi_n(z=0)=0.
\end{eqnarray}
turned to be singular and could not be integrated.
In \cite{Sakai:2004gf}  only the special case of  $u_0=u_{\Lambda}$
was analyzed, in this case since
$\lim _{u_0\to u_{\Lambda}}\partial_u x_{cl}=0$,
a smooth and nonsingular transformation into
cartesian coordinates is allowed via
\begin{equation}\label{ss_tr}
u^3=u_{\Lambda }^3+u_{\Lambda }^3(z^2+y^2)\;\;\;;\;\;\;
x_4=R\arctan(\frac{y}{z})
\end{equation}
The corresponding action for $y$ is (after setting $u_{\Lambda}=1$)
\begin{eqnarray}\label{original_action}
S\sim \int d^4xdz\bigg[ \frac{(\partial_{\mu}y)^2}{u(z)}+u(z)^3(\partial_z y)^2+2y^2\bigg]
\end{eqnarray}
inserting the expantion $y=\sum_{n=1}\varphi_n(x^{\mu})y_n(z)$ the e.o.m for $y_n$ is
\begin{equation}\label{SS_eq}
\partial_z^2y_n+\frac{2z}{1+z^2}\partial_zy_n-\frac{2y_n}{1+z^2}=\frac {m_n^2}{(1+z^2)^{4/3}}y_n
\end{equation}
which is non-singular.
For the more general case of $u_0>u_{\Lambda}$ we were not able to find
a similar
 coordinate transformation and hence we follow a different approach desribed in the next section.

\section{A regular e.o.m for the scalar fluctuation at the low temperature phase}\label{section3}

\begin{figure}
\centerline{
\begin{tabular}{cc}
\includegraphics[width=3in]{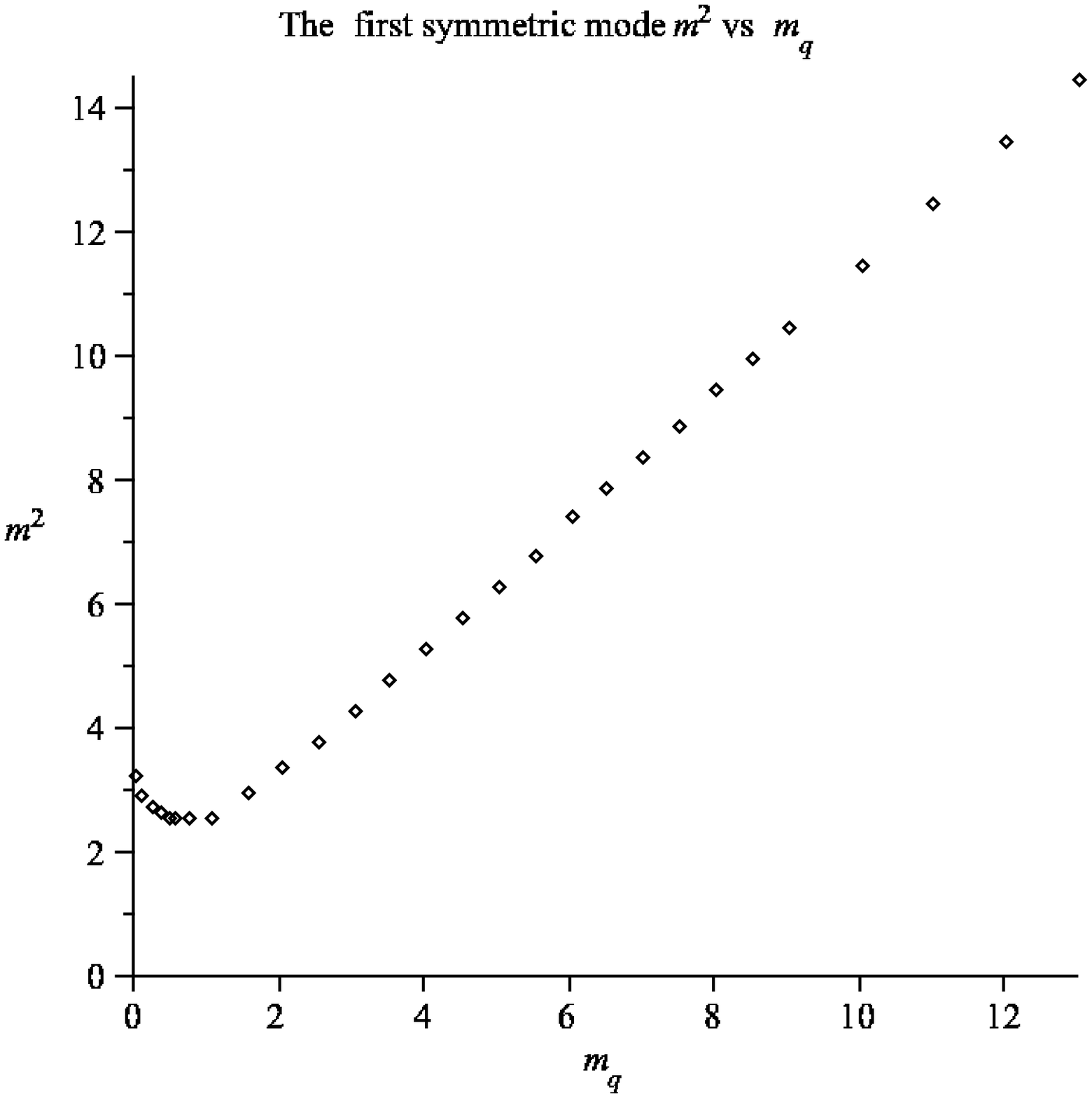}&
\includegraphics[width=3in]{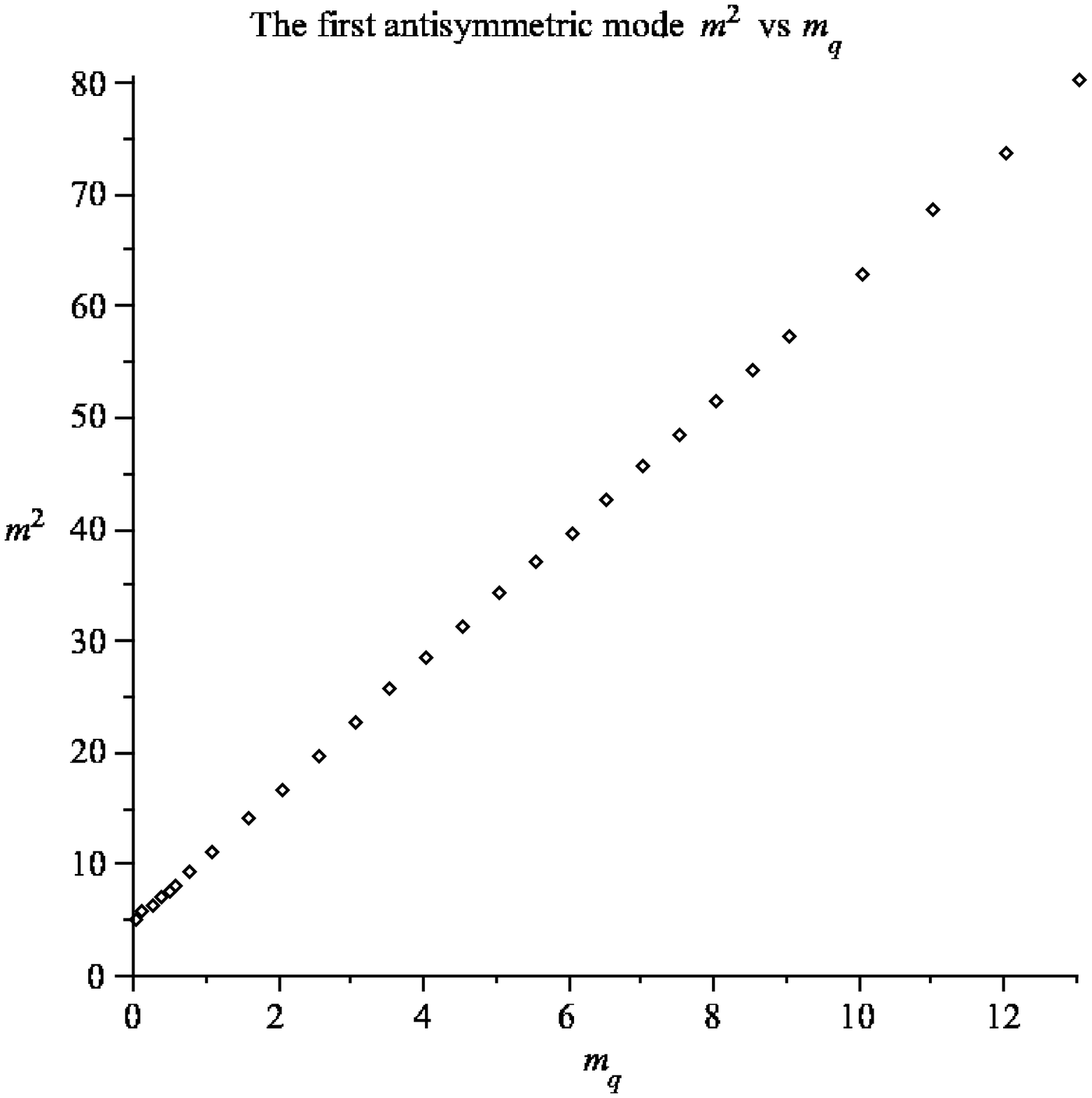}
\\(A)&(B)
\end{tabular}
}
\caption{(A) The mass squared $m_1^2$ of the lowest exited symmetric mode as a function
of $m^c_q$ ($R_{D4}=u_\Lambda=1$) }
\label{m_1}
\caption{(B) The mass squared $m_2^2$ of the lowest exited antisymmetric mode as a function
of $m^c_q$ ($R_{D4}=u_\Lambda=1$)}
\label{m_2}
\end{figure}

\begin{figure}
\centerline{
\includegraphics[width=3in]{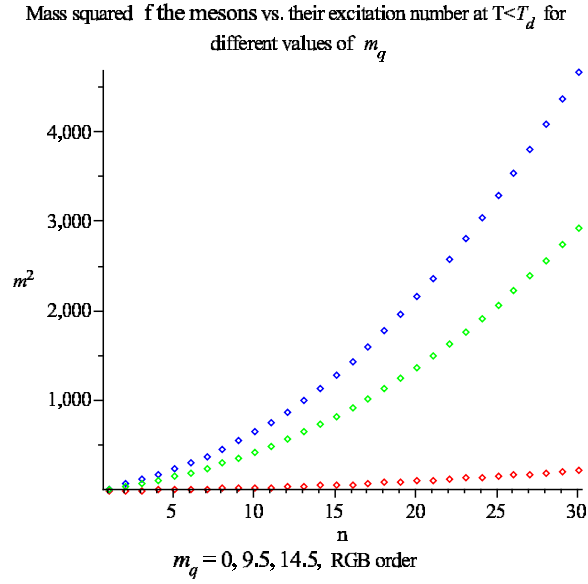}}
\caption{(A) The tower of the mesons squared mass $m^2_n$ in the low temperature phase ($R_{D4}=u_T=1$)}
\label{m_n}
\end{figure}

As we have seen above, we could not
obtain the even modes of the fluctuation \footnote{If the classical curve $x_{4,cl}$ was odd,
then the odd mode would become singular.} around the classical curve
because  $\frac{dx_{4,cl}}{du}$ diverges at $u=u_0$.
The issue of choosing a direction along which one should anlayze the fluctuations, has been discussed in the context of the stringy description of the Wilson like \cite{Kinar:1999xu}. It was found
that the safest approach is to use the fluctuation in the direction which is normal to the classical configuration. For our case the normal to the classical configuration at the tip $u=u_0$ is along the $u$ direction. Thus from here on
 we study the fluctuation in the $u$ direction, that is
\begin{eqnarray}\label{u_u_x}
u(x_4,x^{\mu})=u_{cl}(x_4)+\xi(x_4,x^{\mu})
\end{eqnarray}
our classical curve would be $u_{cl}(x_4)$ and as can be seen from (\ref{partial_x_4}) we have $\frac{du_{cl}}{dx_4}|_{x=0}=0$
so the point $u(x_4=0)=u_0$ pause no problem now!
The quadratic action for these fluctuation is (after setting  $u_{\Lambda}=1$)
\begin{eqnarray}
S&=&\frac{1}{2}\int dx_4\bigg{\{}\frac{a_0}{u^{11}f^3}(\partial_{x_4}\xi)^2+\frac{1}{u^3f}(\partial_{\mu}\xi)^2
\\ \nonumber
&-&\frac{(11u^{14}+18a_0+3u^{11}-12u^8-27a_0(u^3+u^6)-2u^5)}{2u^{16}f^3}\xi^2\bigg{\}}
\end{eqnarray}
where $a_0=u_0^8f(u_0)$ and it should be understood that $u=u_{cl}(x_4)$ and its formal expression is
\begin{eqnarray}\label{u_of_x}
u(x_4)=\int^{x_4}_{0}
dx_4f(u)(\frac{u}{R_{D4}})^{3/2}\sqrt{\frac{f(u)u^8}{f(u_0)u_0^8}-1}
\end{eqnarray}
after plugging a mode expansion
the e.o.m in its canonical form is
\begin{eqnarray}\label{eom_u}
\partial^2_x\xi_n&-&(\frac{11}{u}+\frac{9}{uf})u_x\partial_x\xi_n
-\frac{f^2u^8m_n^2}{a_0}\xi_n \\ \nonumber
&+&\frac{(11u^{14}+18a_0+6u^{11}-12u^8-27a_0(u^3+u^6)-2u^5)}{2a_0u^5}\xi_n=0
\end{eqnarray}
where $u_x=\partial_{x_4}u_{cl}$. Since there is no analytic expression for the integral
in (\ref{u_of_x}), we obtained $u(x_4)$ numerically during the integration of
eq. (\ref{eom_u}) when 'shooting' to find the eigenvalues of (\ref{eom_u}).\\
The resulted spectra  are summarized  in figures (\ref{m_1}), (\ref{m_2}) and (\ref{m_n}). The following properties charcterize these spectra
\begin{itemize}
\item
The first observation one can make is that
 for  $u_0=u_{\Lambda}$ our results for the symmetric and anti-symmetric lowest lying states match
those of \cite{Sakai:2004gf}.
\begin{eqnarray}
m_s^2=3.3\ \ \ \ \   ;\  \ \ \ \ m_{as}^2=5.3
\end{eqnarray}
\item
The figures   (\ref{m_1}) and (\ref{m_2}) describe the dependence of the
squared mass of the  first excited symmetric and anti-symmetric states as a function of the ``constituent quark mass''
defined in \cite{Casero:2005se} and \cite{Kruczenski:2004me}, as follows
\begin{eqnarray}\label{constituentmass}
m^c_q=\frac{1}{2\pi\alpha'}\int^{u_0}_{u_{\Lambda}}\sqrt{-g_{tt}g_{uu}}du
=\frac{1}{2\pi\alpha'}\int^{u_0}_{u_{\Lambda}}f^{-1/2}(u)du
\end{eqnarray}
This parameter relates to the constitutent quark mass and not to the current algebra
( QCD) mass, since even when it is turned on the fluctuations that correspond to the
pions are massless. In fact the quantity dual of the constituent quark mass should
associate with $m^c_q$ plus a constant term which is independent of $u_0$
since  already
for $u_0=u_{\Lambda}$ the mesons are massive and hence there is a non-trivial constituent mass.
This assignment is also in accordance with
the semi-classical description of high spin mesons \cite{Kruczenski:2004me} and
their stringy split into two lower mass  mesons  \cite{Peeters:2005fq}.
>From these figures we see that indeed for  for $u_0>u_\Lambda$ the square
of the  mass of the
scalars grows linearly with $m^c_q$. This is to be contrasted with the results found in  \cite{Casero:2005se} for vector mesons of non-critical models where the masss itself is found to be linear with the $m^c_q$ ( see also down in section (5).)
\item
We have also determined the spectrum of the higher excited mesons both the symmetric as well as the anti-symmetric ones. The dependence
 of the squared masses on the excitation number
for various values of $m^c_q$ is drawn  in figure (\ref{m_n}).
The linear fit to this curves are given by
\begin{eqnarray}\label{mass_trajectory_cr}
&m_n^2&=3.3+1.6n^{1.7} \ \ \ \ \ \ \ \ \ m^c_q=0 \\ \nonumber
&m_n^2&=10.5+6.5n^{1.789}\ \  \ \ \ m^c_q=9.3\\ \nonumber
&m_n^2&=15.8+9.5n^{1.818}  \ \  \ \ \ m^c_q=14.3
\end{eqnarray}

Stringy modes are characterized by the well known  $m^2\sim n$ behavior.
We thus see that  the scalar meson spectra that follow from the model of
\cite{Sakai:2004gf} do not correspond to stringy modes. This is of course of no surprise
since it follows from a low energy effective field theory and not from a semi-classical treatment.
\item
Last by not least we see from figure (\ref{m_1}) that the lowest scalar excitation remain non-tachyonic
for all values of $u_0$ which serves as an partial evidence for
 the stability of the Sakai Sugimoto
model.
\end{itemize}

\section{Scalar mesons in a non critical holographic model}\label{non_crit}
We would like now to find the masses of the scalar modes associated
with the fluctuations of the probe brane around the classical
profile in the non-critical gravity background of
\cite{Kuperstein:2004yf}. Using the background
(\ref{noncriticalbackground}) in an effective action that includes
only the DBI. The CS term \be\label{SCS} S_{CS}\sim \int_{D4}C_5
\sim \int_{D4}\frac{u^5}{R_{AdS}^4}\ee would contribute upon a
substitution of the form (\ref{u_u_x}) for $u$ and expanding a linear and
quadratic term to the action, thus affecting the spectrum. However,
it was found in \cite{Mazu:2007tp} that including this CS does not
yield a sensible thermal phase diagram and hence
we discuss separately an effective action that includes only a
DBI action and one with both the DBI and CS terms. We start
first with the former case:
Analyzing the spectrum in a
similar manner to
the analysis of section (3)  we find  that the fluctuations
are subjected  to the  following eigenvalue equation
\begin{eqnarray}
\partial_u(u^4\gamma^{-3/2})\partial_u\xi_n=-\frac{R_{AdS}^4m_n^{\prime2}u^2}
{\gamma^{1/2}}\xi_n
\end{eqnarray}

Like in the critical case, for $u_0>>u_\Lambda$ the qualitative behavior
of $m'_n$ can be seen by changing the variable $u$ into
the dimensionless parameter
$y=\frac{u}{u_0}$. At the limit $u_0>>u_\Lambda$ we find that $f(u)\to 1$
and so
\begin{eqnarray}
\gamma\to \frac{1}{u_0^2(y^2-\frac{1}{y^8})}
\end{eqnarray}
and find that in terms of the dimensionless parameter $y$ the e.o.m is now
\begin{eqnarray}
\partial_y(y^4\gamma^{-3/2})\partial_y\xi_n=-\frac{R_{AdS}^4m_n^{\prime2}y^2}
{u_0^2\gamma^{1/2}}\xi_n
\end{eqnarray}
Since the L.H.S is dimensionless so is the R.H.S and we find
\begin{eqnarray}\label{m_vs_u_0_non_critic}
m_n^{\prime2} \propto \frac{u_0^2}{R_{AdS}^4}
\end{eqnarray}
Note that due to the different background now $L\sim  \frac{R_{AdS}^2}{u_0}$ and hence again
we get that $m_n'\sim \frac{1}{L}$. However in terms of $m^c_q$ the asymptotic behavior is that
$m_n'\sim m^c_q$ and not ${m_n'}^2\sim m^c_q$ as was the case for the mesons of the critical model.

\begin{figure}
\centerline{
\begin{tabular}{cc}
\includegraphics[width=3in]{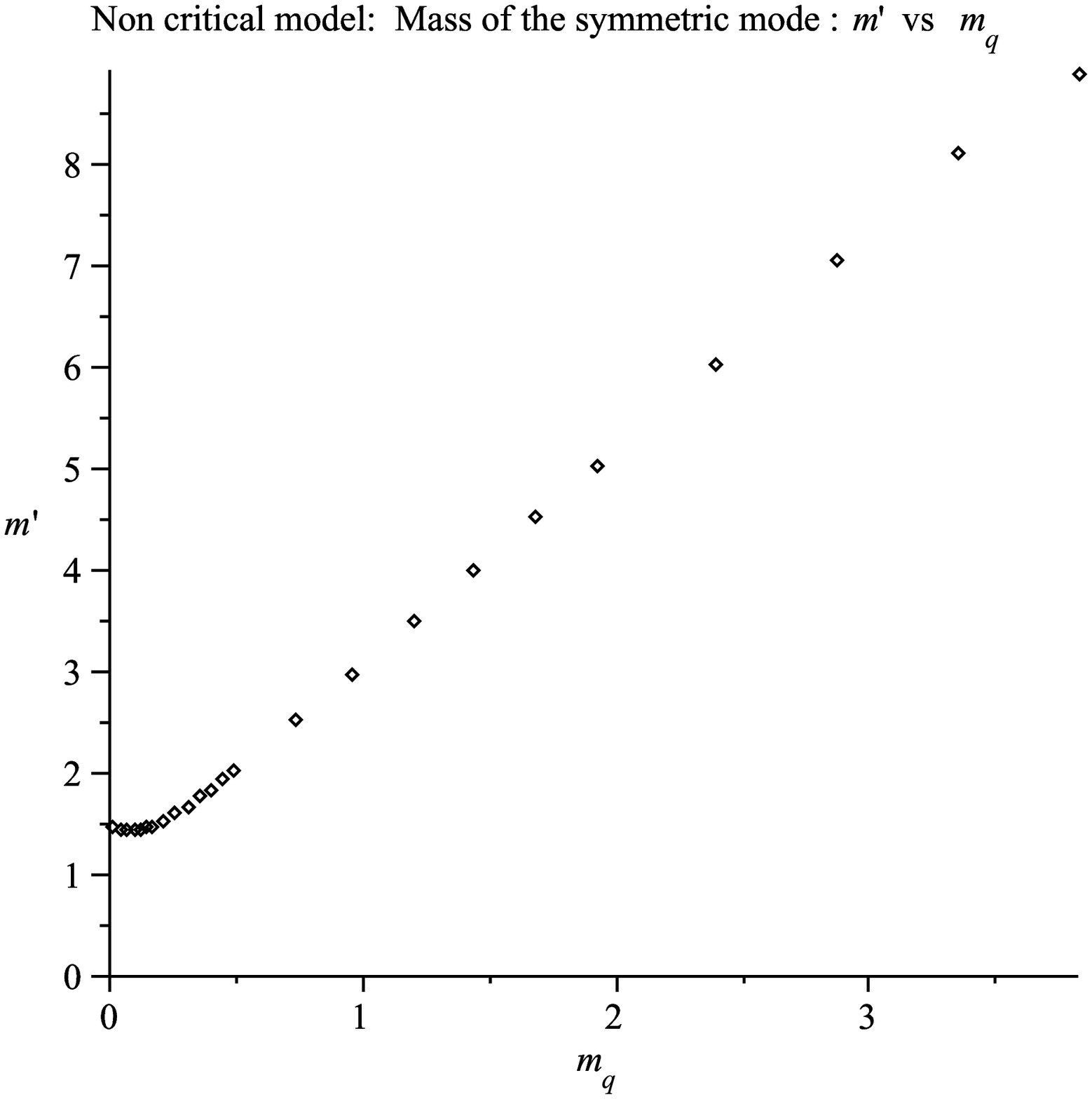}&
\includegraphics[width=3in]{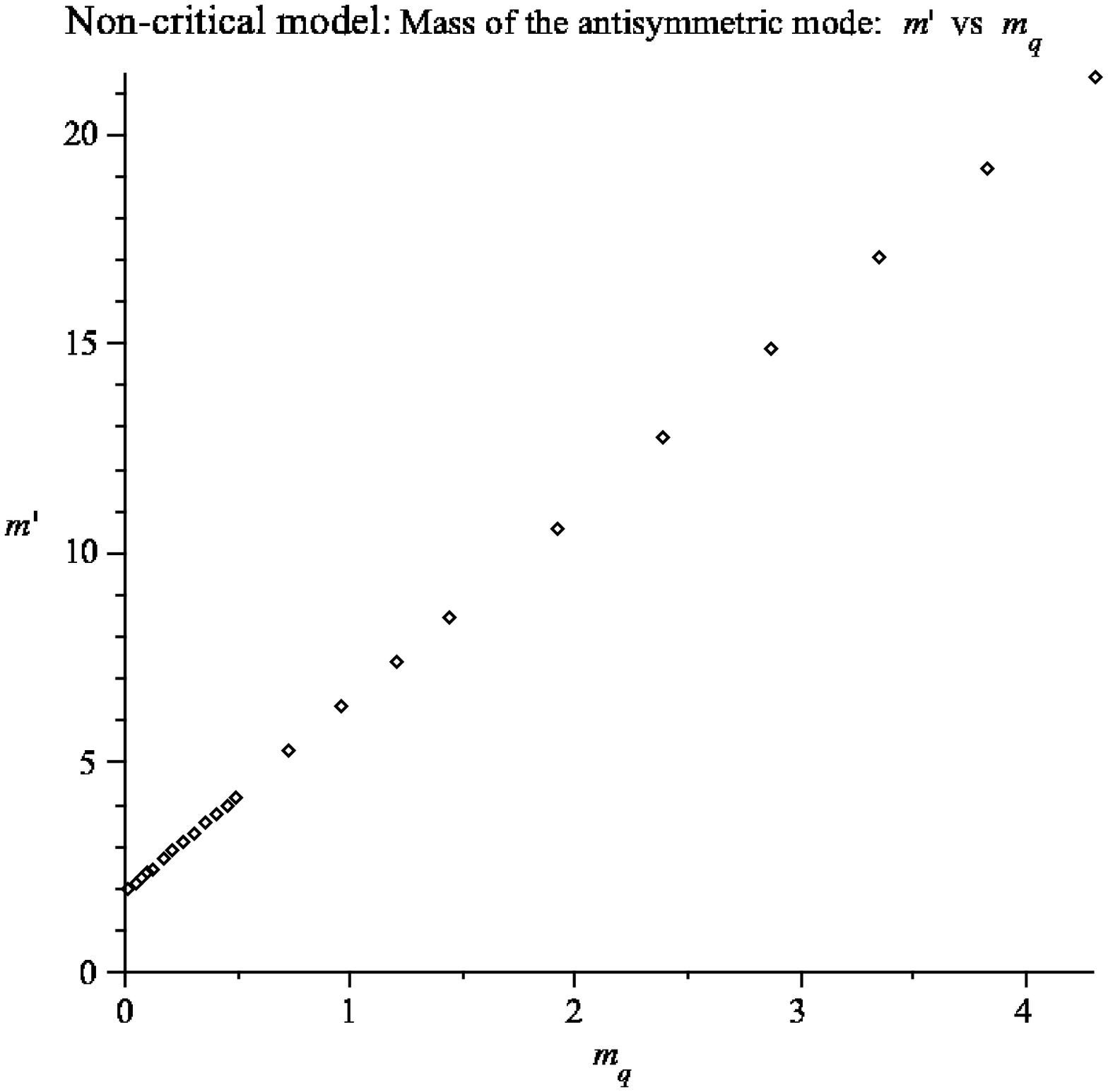}
\\(A)&(B)
\end{tabular}
}
\caption{(A) The mass ${m'}_1$ of the lowest exited symmetric mode
of the non-critical model as a function of $m^c_q$ ($R_{AdS}=u_\Lambda=1$).}
\label{m_non_crit_1}
\caption{(B) The mass ${m'}_2$ of the lowest exited antisymmetric mode
of the non-critical model as a function of $m^c_q$ ($R_{AdS}=u_\Lambda=1$).}
\label{m_non_crit_2}
\end{figure}

\begin{figure}
\includegraphics[width=3in]{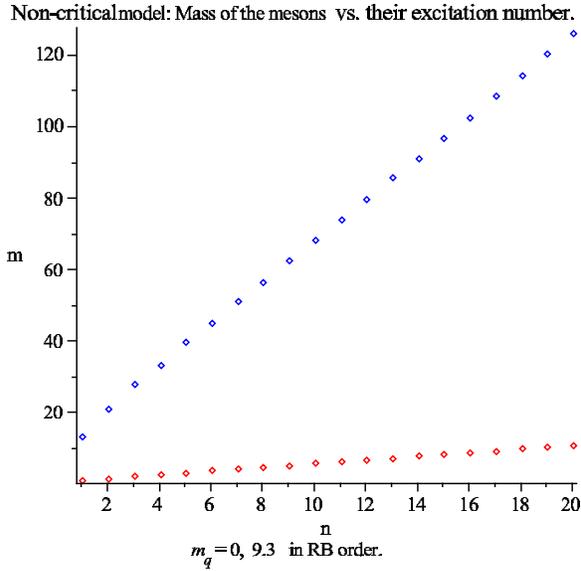}
\caption{The tower of mesons masses $m'_n$ in the non-critical model}
\label{m'_n}
\end{figure}

Repeating the exact same steps as for the critical case we find that the
quadratic action for fluctuation in the $x_4$ direction around the classical
curve leads to an e.o.m which is singular at $u=u_0$ and as a consequence the attempt
carried in \cite{Casero:2005se}
to obtain the spectrum of the even modes had indeed failed.
And so like in the critical case we turn to study the fluctuation in the $u$ direction instead.
The action for the fluctuation is then
\begin{eqnarray}
S&=&\frac{1}{2}\int dx_4\bigg{\{}\frac{a_0^{3/2}}{u^{14}f^3}(\partial_{x_4}\xi)^2+\frac{a_0^{1/2}R_{AdS}^4}{u^4f}(\partial_{\mu}\xi)^2
\\ \nonumber&-&\frac{a_0^{1/2}(u^5+36a_0-63u^{10}+14u^{20}+48u^{15}-92a_0u^5-44a_0^{10})}{2u^{22}f^3}\bigg{\}}
\end{eqnarray}
and indeed this action leads to a regular e.o.m at $u(x_4=0)=u_0$.
\begin{eqnarray}
&&\partial^2_{x}\xi-(\frac{14}{u}+\frac{15}{u^6f})u_x\partial_{x}\xi+
\frac{u^{10}f^2R_{AdS}^4}{a_0}\eta^{\mu\nu}\partial_{\mu}\partial_{\nu}\xi
\cr
&& \cr
&+&\frac{(u^5+36a_0-63u^{10}+14u^{20}+48u^{15}-92a_0u^5-44a_0u^{10})}{2u^8a_0}\ \xi=0
\end{eqnarray}
Using the shooting technic we found the eigenvalues of different
modes of the fluctuation 
for various values of $m^c_q$,
our finding are summarize in figures (\ref{m_non_crit_1}),(\ref{m_non_crit_2}).
One can see that the masses $m'_1$ and $m'_2$ grow linearly with $m^c_q$
as expected from (\ref{m_vs_u_0_non_critic}).
At $u_0=u_{\Lambda}=1$ we find\footnote{Our results are for $R_{AdS}=1$.}.
\begin{eqnarray}
m_s^{\prime2}=1.51\ \ \ \ \   ;\  \ \ \ \ m_{as}^{\prime2}=2.07.
\end{eqnarray}
which is in agrement with \cite{Casero:2005se}. \footnote{To keep contact
with the results in \cite{Casero:2005se} we had renormalized the masses
by the factor $\frac{2}{5}$ coming from the change of variables
$u\to z$.}
Again we also studied the dependence of the mass on the excitation number, the results
are  summarized in figure (\ref{m'_n}) and are:
\begin{eqnarray}\label{mass_trajectory_non_cr}
&m_n&=1.51+ 2.32n^{1.04}\ \ \ \ \ m^c_q=0 \\ \nonumber
&m_n&=13.5+ 4.95n^{1.04}  \ \ \  \  \ m^c_q=9.3
\end{eqnarray}

Thus we see that both in terms of the dependence on $n$ as well as the dependence on
$m^c_q$ the scalar meson spectra admits a different behavior than that of the critical model
of \cite{Sakai:2004gf}. A similar behavior has been observed for the vector mesons in \cite{Casero:2005se}.\\
Next we consider the case where the effective action includes both the DBI and CS terms.
Including now the CS term (with its full strengh $\tilde{a}=1$) the quadratic action for the
fluctuation becomes
\begin{eqnarray}
S&=&\frac{1}{2}\int dx_4\bigg{\{}\frac{B^{3/2}}{u^{14}f^3}(\partial_{x_4}\xi)^2+\frac{B^{1/2}R_{AdS}^4}{u^4f}(\partial_{\mu}\xi)^2
\\ \nonumber&-&\frac{B^{1/2}(u^5+36B-63u^{10}+14u^{20}+48u^{15}-92Bu^5-44B^{10})}
{2u^{22}f^3}-\frac{20}{\sqrt{5}}u^3\xi^2\bigg{\}}
\end{eqnarray}
where $B=(u_0^5f^{1/2}(u_0)-u_0^5+u^5)^2$
and the e.o.m is then
\begin{eqnarray}
&&\partial^2_{x}\xi-(\frac{14}{u}+\frac{15}{u^6f}-\frac{15u^4}
{B^{1/2}})u_x\partial_{x}\xi+
\frac{u^{10}f^2R_{AdS}^4}{B}\eta^{\mu\nu}\partial_{\mu}\partial_{\nu}\xi
\cr
&& \cr
&+&\frac{(u^5+36B-63u^{10}+14u^{20}+48u^{15}-92Bu^5-44Bu^{10})}{2u^8B}\ \xi
+\frac{20u^3}{\sqrt{5}B^{3/2}}\xi=0
\end{eqnarray}

\begin{figure}
\centerline{
\begin{tabular}{cc}
\includegraphics[width=3in]{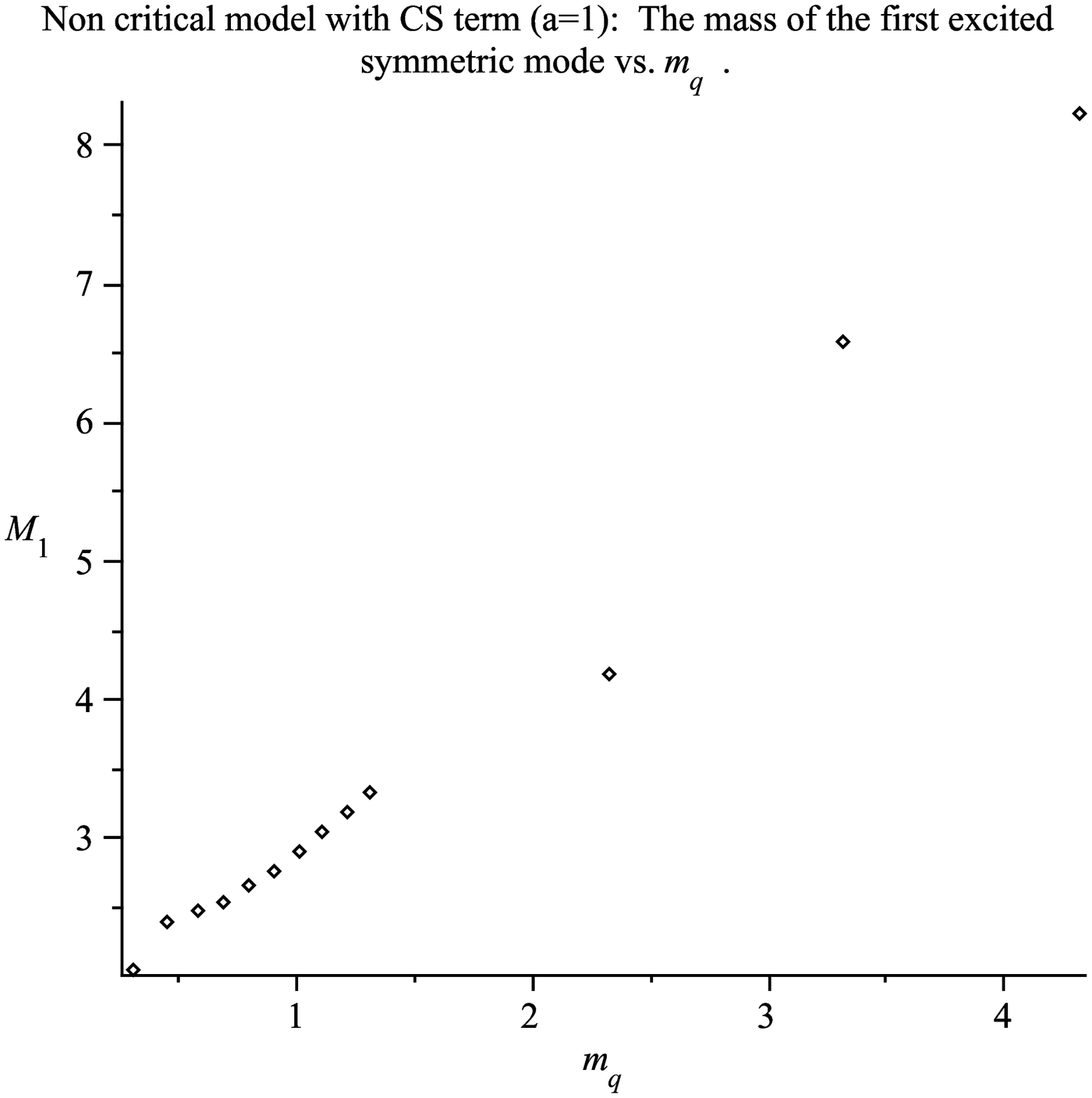}&
\includegraphics[width=3in]{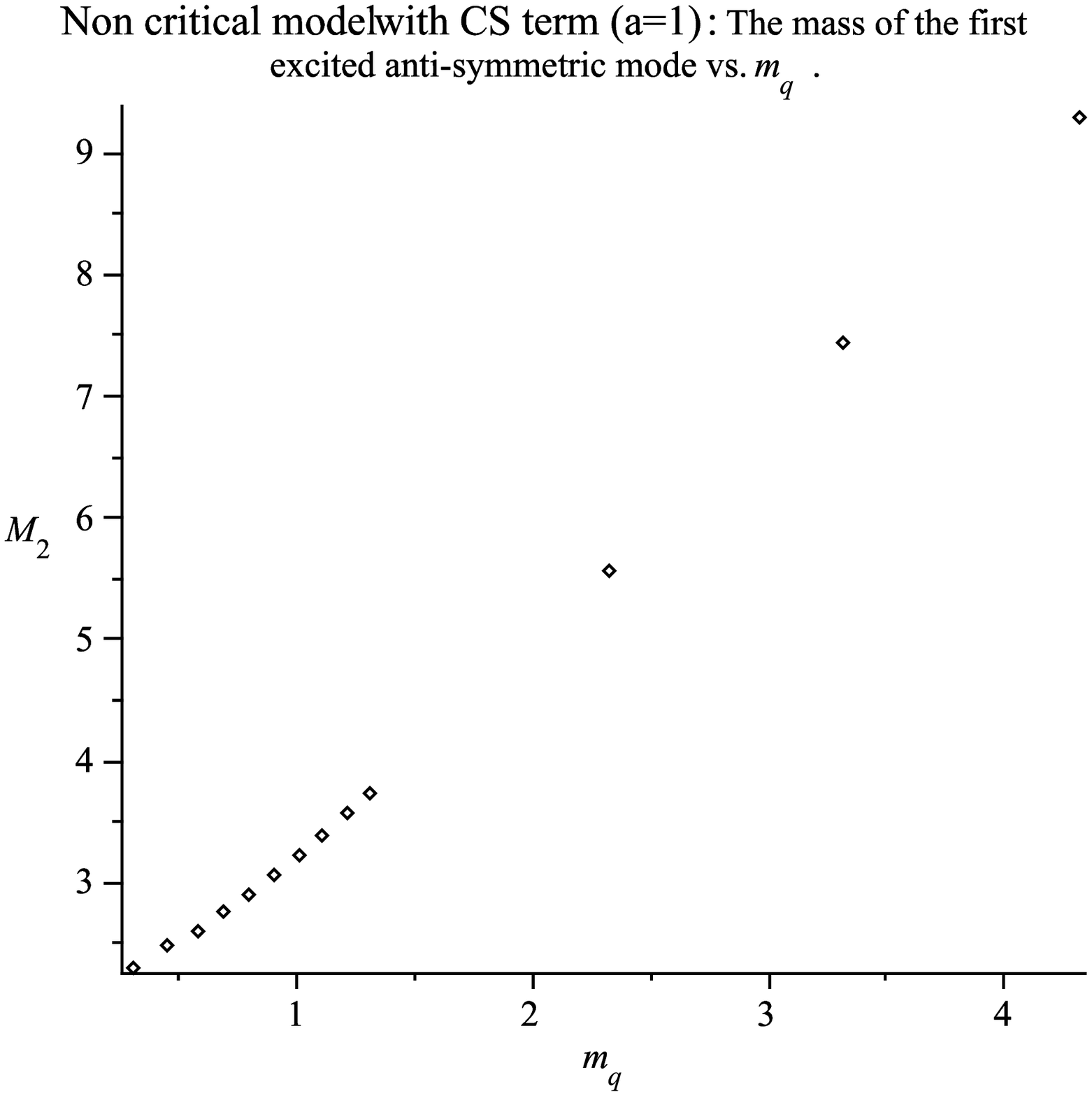}
\\(A)&(B)
\end{tabular}
}
\caption{(A) The mass ${m'}_1$ of the lowest exited symmetric mode
of the non-critical model with CS term included as a function of $m^c_q$ .}
\label{m_non_crit_a_1}
\caption{(B) The mass ${m'}_2$ of the lowest exited antisymmetric mode
of the non-critical model with CS term included as a function of $m^c_q$ .}
\label{m_non_crit_a_2}
\end{figure}

\begin{figure}
\includegraphics[width=3in]{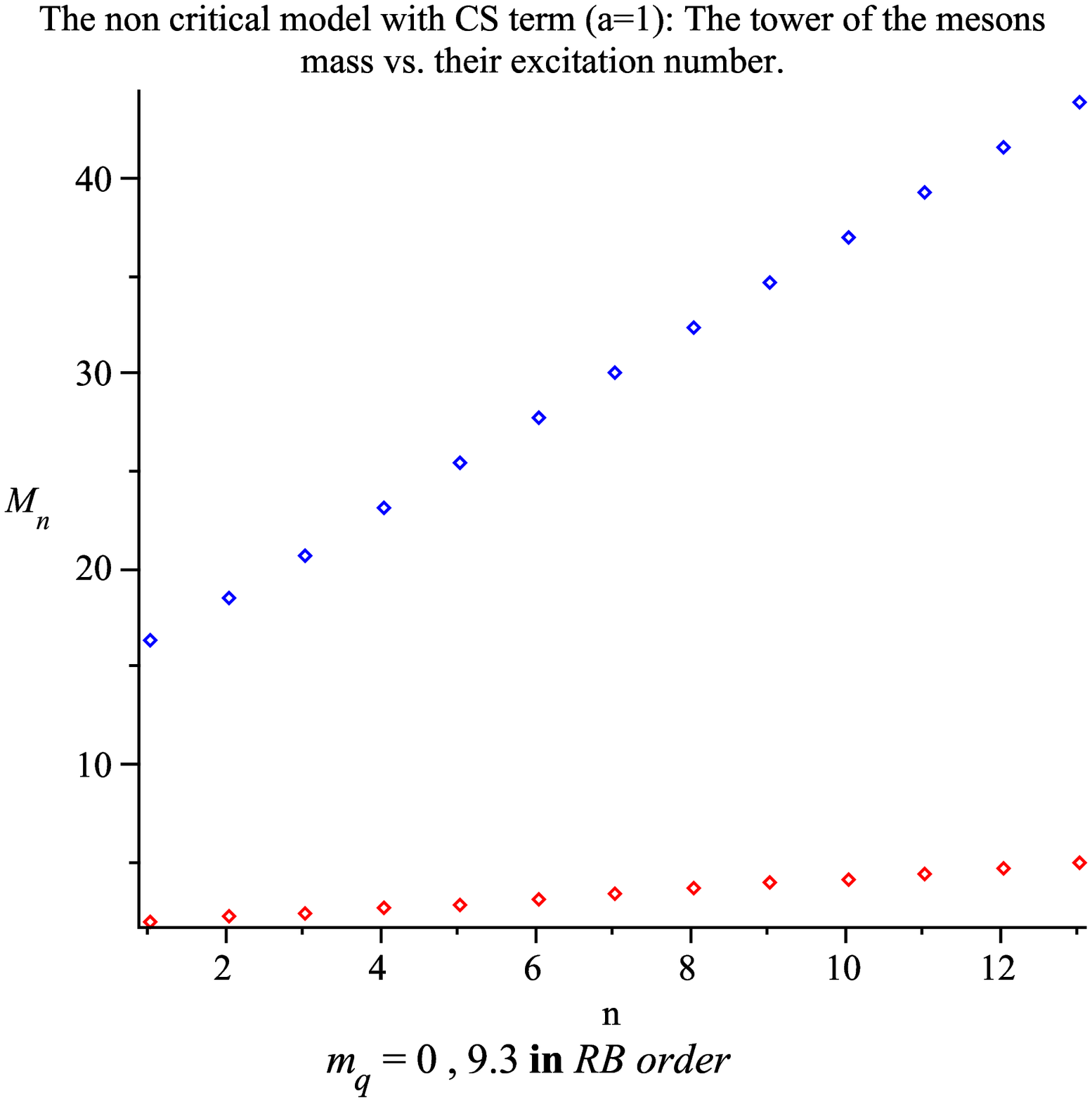}
\caption{The tower of mesons masses $m'_n$ in the non-critical model with CS term included}
\label{m'_n_a}
\end{figure}

With the Chers Simon taking into account the dependence
of the mass squared on the excitation number is now to
be read from figure (\ref{m'_n_a}) to be:
\begin{eqnarray}\label{mass_trajectory_non_cr}
&m_n&=2.07+ 5.42n^{0.75}\ \ \ \ \ m^c_q=0 \\ \nonumber
&m_n&=16.49+ 1.01n^{0.75}  \ \ \  \  \ m^c_q=9.3
\end{eqnarray}
The dependance on $m^c_q$ is described in figures (\ref{m_non_crit_a_1})
and (\ref{m_non_crit_a_2}).
\section {Parity and charge conjucation}\label{parity}
In order to compare the resulting spectra from both the critical and
non-critical models, we first have to identify the ``quantum
numbers'' of the states that correspond to the fluctuations. More
explicitly  we have to determine the operations in the gravity
models which correspond to charge conjugation and parity
transformations. In the model of  \cite{Sakai:2004gf} they were
defined as follows: The {\it  charge conjugation}  operation
associates with exchanging the left and right handed quarks which
maps into the interchange of a $D8$ and an anti $D8$ or differently
transforming $z\rightarrow -z$. {\it Parity} transformation in the
five-dimensional space-time spanned by $x_i,z$ where $i=1,2,3$ means
the following transformation $( x_i, z) \rightarrow (-x_i, -z)$.

For the generalized set up with $u_0>u_\Lambda$ we can still define
the coordinate $z$ as follows \be  u^3 = u_0^3 + u_\Lambda z^2 \ee
Note the difference with respect to (\ref{defz}) since here we take
$z$ to have dimension of length. With  this definition of the $z$
coordinate the discrete transformations of \cite{Sakai:2004gf}
remain in tact. The effective action on the probe brane has to be
invariant under both parity and charge conjugation. The DBI part
(\ref{DBIscalar}) is quadratic in $\xi$ and hence cannot  determine
the right transformation of the fluctuation modes. The situation
with the CS term is different. Recall that the CS term has the form
\be S_{CS}\sim \int_{D8} F\wedge F\wedge C_5 =\int_{S^4}F\wedge
F\wedge \int d^4 x dz C_5 = \int_{S^4}F\wedge F \int d^4 x dz
\xi(x^\mu,z) \ee The  last part we have used the explicit form of
the $C_5$
\be C_5 =\xi(x^\mu,z) dx^0\wedge...dx^3\wedge dz \ee
In order for this term  in the action to be invariant under parity and
charge
 conjugation it is clear that
 $\xi(x^\mu,z)$ has to be even under both charge confugation and
parity transformation. Now since $\xi(x,z) = \sum_n
f_n(x^\mu)\xi_n(z)$ we conclude that the map between the fluctuation
modes and sclar particles is the following \bea\label{PCas}
symmetric \ &\xi_n& \qquad \rightarrow \ \ \ 0^{++}\  mesons \cr
antisymmetic &\xi_n& \qquad \rightarrow\ \ \   0^{--}\  mesons  \eea

For the non-critical model again the DBI action does not determine
the transformations of $\xi$ under parity and charge conjugation. We
have argued above based on \cite{Mazu:2007tp} that a CS term of
the form  (\ref{SCS}) should not be incorporated. Thus there is no
way to this order to determine the transformation of $\xi$.

 Without the constraint from the
 CS term we may have that $\xi$ is even or odd under charge
 conjugation and parity transformtaions. In the latter case the
 assignments of (\ref{PCas}) have to be reversed, namely symmetric
 $\xi$ corresponds to $0^{--}$ and antisymmetric $\xi$ to $0^{++}$

Next we want to compare the spectra to mesons observed in nature.
It is well know that scalar mesons in nature are either $0^{++}$ or pseudo sclars of the form $0^{-+}$ and there are no observed low lying mesons of the form
$0^{--}$.
 Thus there is a serious mismatch between the holographic scalar mesons
 extracted from models with flavor branes anti-branes
 of critical  models  and with the observed mesons in nature.
  We will come back to this issue in the conclusions.

\section{Scalar mesons in the intermediate temperature phase}\label{intermediate}

The background that corresponds to the deconfined phase namely
$T>1/2\pi R$ is given in (\ref{High_T_S_S_metric}). As was shown in \cite{Aharony:2006da}
this deconfined background can admit also a phase where chiral symmetry is broken, the so called ``intemediate phase''
We now analyze the spectrum of the scalar mesons in this phase.
Since the procedure of extracting the scalar meson
is identical to that of the low temperature analysis
of the previous sections we present the final results for the spectra of masses.
The spectra are presented in figures (\ref{mT_1}), (\ref{mT_2}) and ({\ref{m_T_n}). The main features that these spectra admit are the following
\begin{itemize}
\item

As can be seen, at the phase transition $T=T_d$ the value of the
masses are (for the values $u_T=1,u_0=8$)
\begin{eqnarray}
m_s^2 (T=T_d)=8.36\ \ \ \ \   ;\  \ \ \ \ m_{as}^2(T=T_d)=45.96
\end{eqnarray}
while in the low temperature phase at the point of phase transition with
$u_{\Lambda}\to u_T=1$, $u_0=8$  the masses are
\begin{eqnarray}
m_s^2(T=T_d)=8.40\ \ \ \ \   ;\  \ \ \ \ m_{as}^2(T=T_d)=46.00
\end{eqnarray}
We see a very small jump in the masses at the transition point, the
same as was seen for the vectors in \cite{Peeters:2006iu}
\item
While in the low temperature confining phase the masses of the mesons are
temperature independent since the background in this phase does not depend on the temperature, the masses of the mesons do depend on the temperture in the
intermediate deconfined phase. As was observed in Lattice simulations and was
found also for holographic vector mesons \cite{Peeters:2006iu}, the masses decrease as a function of the temperature. The symmetric mesons decrease at the chiral symmetry phase transition temperature   $T=T_{\chi SB}$ to $\sim 60\%$ persent of their values whereas the antisymmetric ones to $\sim 80\%$. This drop off is much more significant than for the vecotr mesons of the critical model \cite{Peeters:2006iu}.

\item
 Note that it is
only consistent to increase the temperature up to where the next
phase transition occur and chiral symmetry is restored.\\
This happens at $T=T_{\chi SB}$ (for the choice $u_0=8$ we found that $T_{\chi SB}=2.44T_d$),
then the merged together   $D8$-$\bar{D}8$ breaks into a separate pair of $D8$-$\bar{D}8$.
We can also see from figure (\ref{mT_1}) that if we continue to
increase the temperature beyond $T_{\chi SB}$
then at some point the scalar mode becomes Tachyonic, signaling that
this background is no longer stable at this temperature as indeed we know.\\
\item
Like in the low temperature we had also checked the squared masses dependence
on the excitation number see figure (\ref{m_T_n}), this was found to be:
\begin{eqnarray}
&m_n^2&=8.3+6.4n^{1.7}\ \ \ \ \ T=T_d \\ \nonumber
&m_n^2&=7.6+6.9n^{1.65}\  \ \ \  \ T=2T_d
\end{eqnarray}
\end{itemize}
\begin{figure}
\centerline{
\begin{tabular}{cc}
\includegraphics[width=3in]{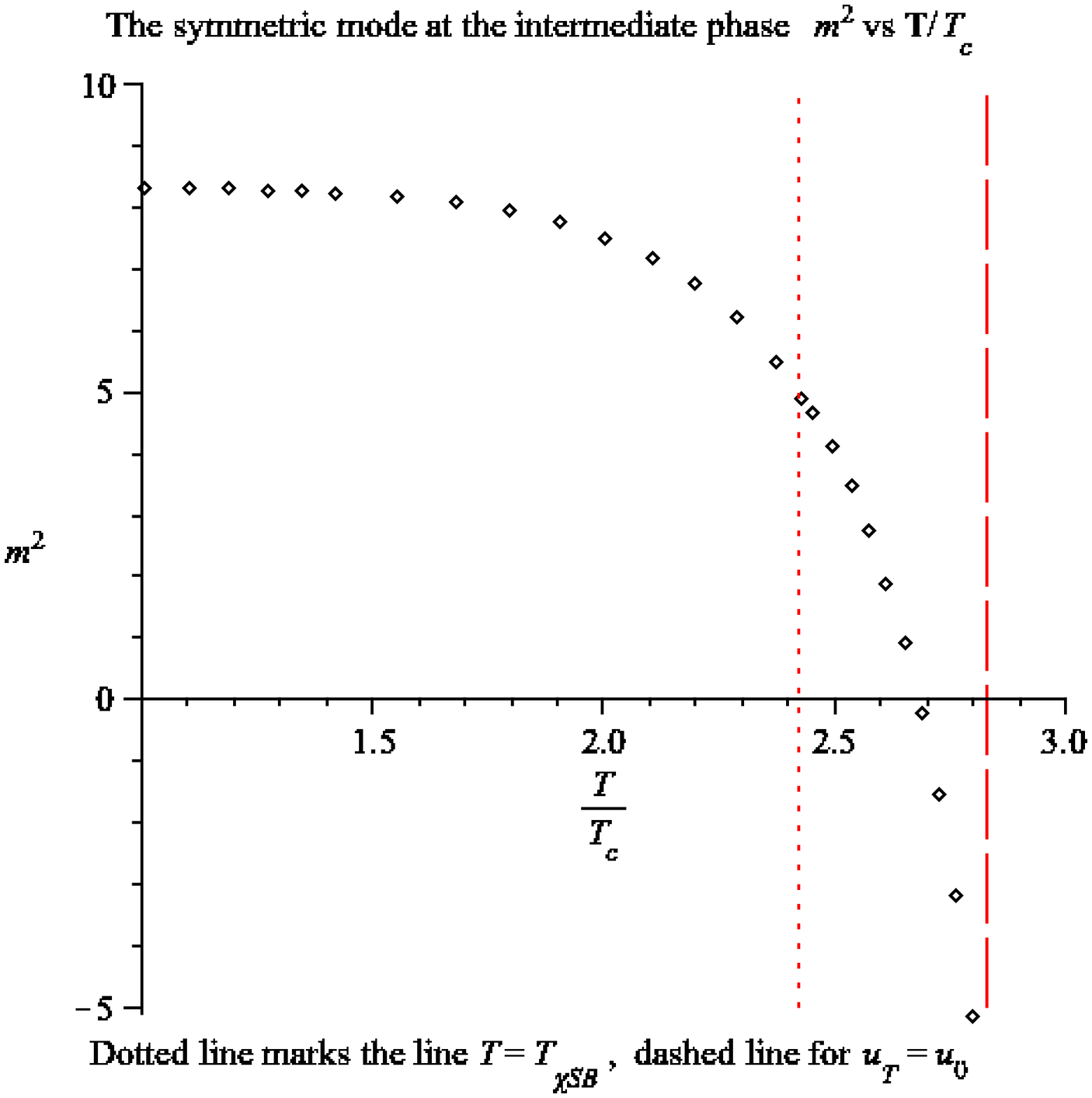}&
\includegraphics[width=3in]{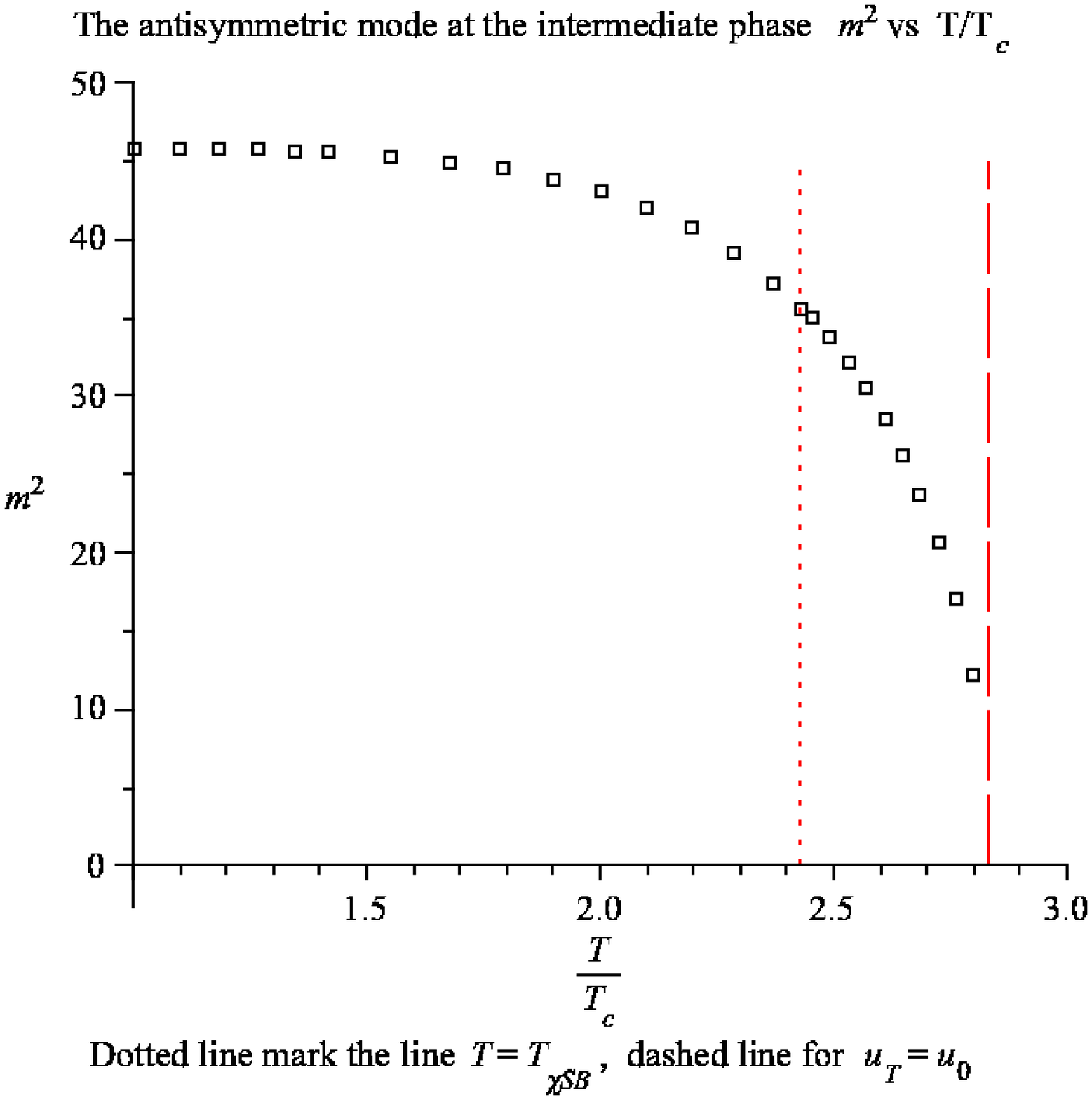}
\\(A)&(B)
\end{tabular}
}
\caption{(A) The mass squared $m_1^2(T)$ of the lowest exited symmetric mode as a function of $T/T_d$
 ($u_0=8,R_{D4}=1$ and $R=2/3$) }
\label{mT_1}
\caption{(B) The mass squared $m^2_2(T)$ of the lowest exited antisymmetric mode as a function of $T/T_d$
 ($u_0=8,R_{D4}=1$ and $R=2/3$)}
\label{mT_2}
\end{figure}

\begin{figure}
\includegraphics[width=3in]{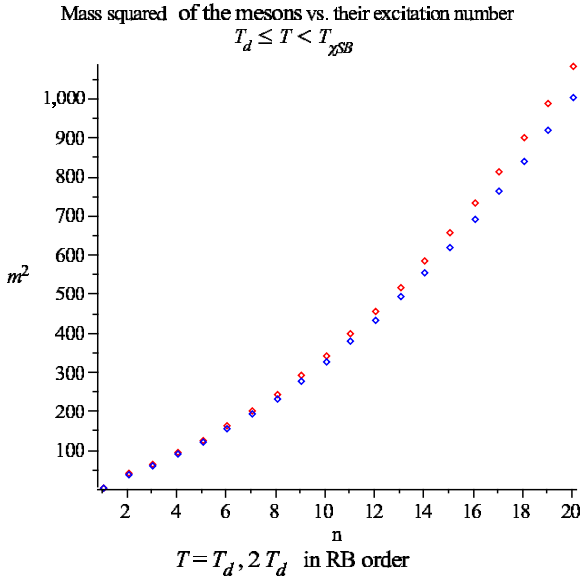}
\caption{The tower of mesons squared mass $m_n^2$ in the intermediate phase}
\label{m_T_n}
\end{figure}

\section{Conclusions}\label{section4}
I this paper we had overcome technical problems  faced in
\cite{Peeters:2005fq,Casero:2005se} and
succeeded to obtain the holographic mass spectra of the scalars in
the low and intermediate phases of the chiral symmetry broken phase
of the critical model and also of those of the non-critical. Let us
summarize the results of this work and mention certain open
directions.
\begin{itemize}
\item
There is a difference between the dependence of the mass of the
scalar mesons on the ``constituent mass parameter" $m^c_q$. In the ten
dimensional models one finds a $m^2\propto m^c_q$ relation (see figures
(\ref{m_1}),(\ref{m_2}) for the first two excited modes), whereas
for the non-critical model the relation is $m\propto m^c_q$ (see
figures (\ref{m_non_crit_1}),(\ref{m_non_crit_2}) and
(\ref{m_non_crit_a_1}),(\ref{m_non_crit_a_2})).
\item
Both the critical models and the non-critical one do not admit a
Regge/stringy behavior of $M_n^2\sim n$. This is not unexpected
since the stringy excitations is not visible in the low energy
effective field thoery.
\item
One can compare the ratio of the low lying mesons both vector and
scalar mesons to those observed in nature. Table (1) present such a
comparison.
\begin{table}
\caption{A comparison to the experimental data where the best fitted
$m^c_q$ is presented vs. $m^c_q=0$ (for the critical case we
have found that there is no improvement in ratios of the vectors so we
have left these entries empty.).}\label{exp_data}
\begin{tabular}{|c|c|c|c|}
\hline \ \  & experiment & D4-D8 at  $m^c_q=0\ /\ 0.38$  &
Non-critical at $m^c_q=0\ /\ 0.16$ \\
\hline
$m_{v,2}^2/m_{v,1}^2$  &2.51 & 2.4\ /\ - & 2.8\ /\ 2.62 \ \\
\hline
$m_{v,3}^2/m_{v,1}^2 $ &3.56 & 4.3\ /\ - & 5.5\ /\ 5.29 \\
\hline
$m_{s}^2/m_{v,1}^2$  &3.61 & 4.9\ /\ 3.63 & 4.1\ /\ 3.65 \\
\hline
$m_{v,2}^2/m_{s}^2 $ &0.7 & 0.49\ /\ 0.62 & 0.67\ /\ 0.75 \\
\hline
\end{tabular}
\end{table}
It is interesting to note that turning on a constituent mass $m^c_q$
improves the ratios with respect to those for zero $m^c_q$.
\item
The hologrphic spectra of the critical models admit a branch of scalar
mesons of the type $0^{--}$. These does not exist in nature. It seems
to be a severe shortcoming of these holographic models. This
difference cannot be attributed to the fact that we consider large
$N_c$. It will be  interesting to investigate the question of how
generic is this situation and whether one can construct a mechanism
to project it out from the low lying spectra.
\item
The behavior of the scalar mesons at finite temperature in the
intermediate phase is similar to that for the vector meson in the
model of \cite{Peeters:2006iu}. However the decrease of the mass with
increasing temperature is more dramatic for the scalar mesons. It is
interesting to check if a similar phenomenon occurs also in lattice
simulations.
\end{itemize}
\section*{Acknowledgments}
We would like to thank Kasper Peeters, Tadakatsu Sakai and Marija Zamaklar for useful discussions,
and specially to
Ofer Aharony for many insightful conversations.
This work was  supported in part by a center of excellence
supported by the Israel Science Foundation (grant number 1468/06), by
a grant (DIP H52) of the German Israel Project Cooperation, by a BSF grant  and by the
European Network MRTN-CT-2004-512194

\bibliographystyle{unsrt}
\bibliography{hscalarsfin}

\end{document}